\documentclass[10pt]{article}

\usepackage{amsmath}
\usepackage{amssymb} 
\usepackage{caption}
\usepackage{graphicx}
\usepackage{authblk}
\usepackage[hypertexnames=false,colorlinks=true,linkcolor=blue,citecolor=blue]{hyperref}
\usepackage[numbers,comma,square,sort&compress]{natbib}
\usepackage[letterpaper,text={6.5in,9in},centering]{geometry}

\graphicspath{{eps/}{pdf/}}
\makeatletter\@addtoreset{equation}{section}\makeatother

\newcommand{\Pc}{\mbox{$P^{\rm C}$}}

\newcommand{\D}{\mathrm{d}}
\newcommand{\E}{\mathrm{e}}
\newcommand{\im}{\mathrm{i}}
\newcommand{\cL}{\mathcal{L}}
\def\sgn{\mathop{\rm sgn}\nolimits}

\begin{document}

\title{On the interaction of uni-directional and bi-directional
  buckling of a plate supported by an elastic foundation}

\author[1]{M. Khurram Wadee}
\author[2]{David J.B. Lloyd}
\author[3]{Andrew P. Bassom}

\affil[1]{\small College of Engineering, Mathematics and Physical
  Sciences, University of Exeter, Exeter, EX4 8EL, UK}
\affil[2]{\small Department of Mathematics, University of Surrey,
  Guildford, GU2 7XH, UK}
\affil[3]{\small School of Mathematics and Physics, University of Tasmania,
  Private Bag 37, Hobart 7001, Australia.}

%\subject{Structural mechanics, nonlinear dynamics, bifurcation
%  analysis, elastic stability}

%\keywords{Numerical solutions, bifurcation analysis, post-buckling}

%\corres{M. Khurram Wadee \email{m.k.wadee@ex.ac.uk}}

\maketitle

\begin{abstract}
  A thin flat rectangular plate supported on its edges and subjected
  to in-plane loading exhibits stable post-buckling behaviour.
  However, the introduction of a nonlinear (softening) elastic
  foundation may cause the response to become unstable.  Here the
  post-buckling of such a structure is investigated and several
  important phenomena are identified, including the transition of
  patterns from stripes to spots and back again.  The interaction
  between these forms is of importance for understanding the possible
  post-buckling behaviours of this structural system.  In addition,
  both periodic and some localized responses are found to exist as the
  dimensions of the plate are increased and this becomes relevant when
  the characteristic wavelengths of the buckle pattern are small
  compared to the size of the plate.  Potential application of the
  model range from macroscopic industrial manufacturing of structural
  elements to the understanding of micro- and nano-scale deformations
  in materials. % which could help us to understand naturally observed
  % phenomena or to use the patters to synthesize materials with
  % specifically tuned physical properties.
\end{abstract}

\section{Introduction}
Instability in elastic bodies is of practical importance to engineers
who are responsible for the design of safe structures. If our sole
concern was to understand the initial stages of buckling of such
bodies then a suitable model consisting of linearized equations would
serve the purpose. On the other hand a deeper appreciation of the true
nature of the buckling together with knowledge of the properties of
later stages of the failure mechanisms requires that nonlinear effects
be included as well.  In the overwhelming majority of post-buckling
scenarios exact analytical solutions do not exist and so, necessarily,
we have to resort to approximation, be it fully numerical or a
simplification of the full system followed by analysis in various
limited regimes \cite{TH73}.  Both of these strategies can reveal
behaviour which is of interest not only to engineers but to applied
scientists and mathematicians as well.  Instability phenomena are
becoming of increasing importance in many fields \cite{CastGom12} and
some previous studies are being given renewed prominence owing to the
discovery of physical phenomena down to the atomic level which these
theories seem to model well \cite{Lee08}.  It is in this spirit that
here we study a problem that at first sight seems to be a simplistic
two-dimensional extension of an exhaustively-studied one, namely the
buckling of a strut resting on an elastic foundation.

In structural engineering, plates are used as load-bearing elements in
various guises, including floor slabs, retaining walls, pad
foundations, ship hulls and aircraft wings.  Of these, perhaps the pad
would most naturally be modelled by a plate on a foundation.  There
are examples where certain parts of structural components can be
modelled as plates which are supported; an obvious candidate is a
sandwich panel for which a core supports a pair of thin skins.
Several of these problems have been examined assuming one-dimensional
behaviour and our intention is to extend these ideas to incorporate a
second dimension. Moreover the importance of supported plates is not
limited to large-scale structures for similar systems have been
proposed as models to look at thin elastic films supported by a
substrate.  There has been much work on one-dimensional
\emph{wrinkling} (another form of buckling) either when a specimen is
compressed or if the supporting substrate (foundation) is
pre-stretched when the elastic layer is placed on it and is then
released.  The plate equations for such a configuration have been
solved using a finite-element technique and compared with experimental
data \cite{Brau}.  Moreover, rather than just being an unfortunate
by-product of loading, wrinkling in some physical systems may actually
be desirable and its control can be effected by judicious application
of load in various directions \cite{CaoW1}.

At an even smaller scale, studies have shown that wrinkling of
nano-scale materials can be sufficient to modify electrical, optical
and other physical properties.  In particular compressed graphene
behaves in this way and a useful model for this material is that of an
isolated elastic layer \cite{GrapheneBuck}.  We believe that improved
understanding of wrinkling in nano-materials might be gained by
incorporating a supporting elastic medium as an integral part of the
system.  %The focus in this field is
%on being able to tune required properties using various manufacturing
%techniques where recent work proposes using loading to influence to
%induce wrinkling \cite{Yang15}.

The well-studied one-dimensional strut model has numerous applications
from the buckling of submarine pipelines and railway tracks through to
sandwich panels and as a phenomenological model for various shell
structures \cite{ChHT,PelTr,JMTTShells}.  Various mathematical
techniques have been utilized to study this system including
Rayleigh--Ritz and Galerkin analyses \cite{WHW}, multiple-scale
perturbation expansions \cite{HBT,HW} and advanced hyperasymptotic
methods \cite{WaBa,ChKo}.  Part of the motivation for this flurry of
activity has been the fact that the strut model can be regarded as a
member of a family of problems governed by a Swift--Hohenberg (SH)
equation.  The SH equation is a partial differential equation in both
time and space that simplifies to an ordinary differential equation
(ODE) form when interest is restricted to time-invariant behaviour in
one spatial dimension.  Static equilibria and their stability can
readily be deduced from this reduced problem.  A key distinction
between the strut and plate models is that the former is of unbounded
length---suitable for the study of very long structures---and
solutions sought can be periodic or homoclinic (localized in extent)
where deflection approaches zero in both positive and negative
directions \cite{Vir}.  In reality of course, at least at the
macroscopic scale, plates are finite and the choice of boundary
conditions potentially have a greater effect on the response of the
system to load than if they are modelled as being completely remote
from the location of the buckle pattern.  Nevertheless, if situations
arise for which the deflection is restricted to a small portion of the
structure then we could say such buckle patterns are effectively
localized in an engineering sense although not homoclinic according
to a strict mathematical definition.

In order to make headway with our stated goal of introducing a second
spatial dimension, it may seem natural to simply investigate the
appropriate planar version of the SH equation; see for
example~\cite{lloyd2008,avitabile2010} for several investigations into
2D localized patterns possible with the planar SH.  However, extending
in a simple-minded manner hides the complication that a flat plate may
respond in other ways when loaded.  For instance, there is now the
possibility of bending about two axes and twisting which gives the
plate extra stiffness and hence offers a good solution for efficient
design \cite{TimWo}.  Furthermore, compatibility and the constitutive
equations arising from the theory of elasticity need to be satisfied
and this adds extra terms together with the need to introduce a new
stress function which relates deformations and strains to stresses.
For moderate to large deflections the bending of a flat plate is often
modelled by the F\"oppl--von~K\'arm\'an (FvK) equations for, although
it is known that they have some limitations and inconsistencies, they
have been shown to usefully model post-buckling plate behaviour at a
qualitative level that is normally accurate enough for most
engineering purposes \cite{Ciarlet}.

We aim to study here the case where the behaviour of an elastic plate
is modified owing to the presence of a supporting elastic foundation
or substrate on which it rests. Depending on the detailed properties
of the foundation different phenomena can be realized.  We will
discuss the various different buckling modes that are possible and how
they relate to each other as the loading on the plate is varied.  In
order to demonstrate the potential usefulness of the plate model, we
will mention a few examples from various fields that might benefit
from our findings.

An important feature of the plate model is that it necessarily
involves separate loading parameters from two directions. For the
one-dimensional strut any loading acts at a point and so cannot vary
spatially but in two directions this constraint has to be relaxed; the
loading in either direction need not be constant and, in general,
would vary with position.  Moreover proper consideration has to be
taken of the aspect ratio of the plate. One might na\"{\i}vely expect
that a long thin plate might well respond to loading in a way similar
to that of a strut but if the two dimensions of the plate are
comparable then it is far from clear what one might expect to
happen. It would not be surprising if in this case distinct phenomena
occur that are peculiar to the strictly two-dimensional case.

The remainder of the paper is organized as follows.  In
\S\ref{sec:modform}, the model is formally introduced and in
\S\ref{sec:galerkin} a Galerkin method is used to obtain approximate
buckling solutions which emerge from the flat fundamental state. These
Galerkin predictions suggest how to conduct efficient full numerical
solutions.  A dynamical stability analysis described in
\S\ref{sec:stab} establishes criteria that predict when a
one-dimensional (striped) solution may be unstable to transverse
instability leading to a fully two-dimensional pattern.  Fully
numerical solutions are obtained and described in \S\ref{sec:num} and
these are compared directly with the earlier results.  The paper
closes in \S\ref{sec:conc} with some conclusions of the work and a
look forward to possible further investigations of the problem.

\section{Model formulation}
\label{sec:modform}
The FvK equations for plates \cite{TimWo} have been shown to provide
an accurate model for post-buckling behaviour and an analysis
demonstrates that the naturally emerging post-buckling solutions are
periodic in two orthogonal directions with the formation of a
rectangular cellular pattern of undulations \cite{EvHu99}.

\begin{figure}
  \centering
  \includegraphics[width=0.8\textwidth]{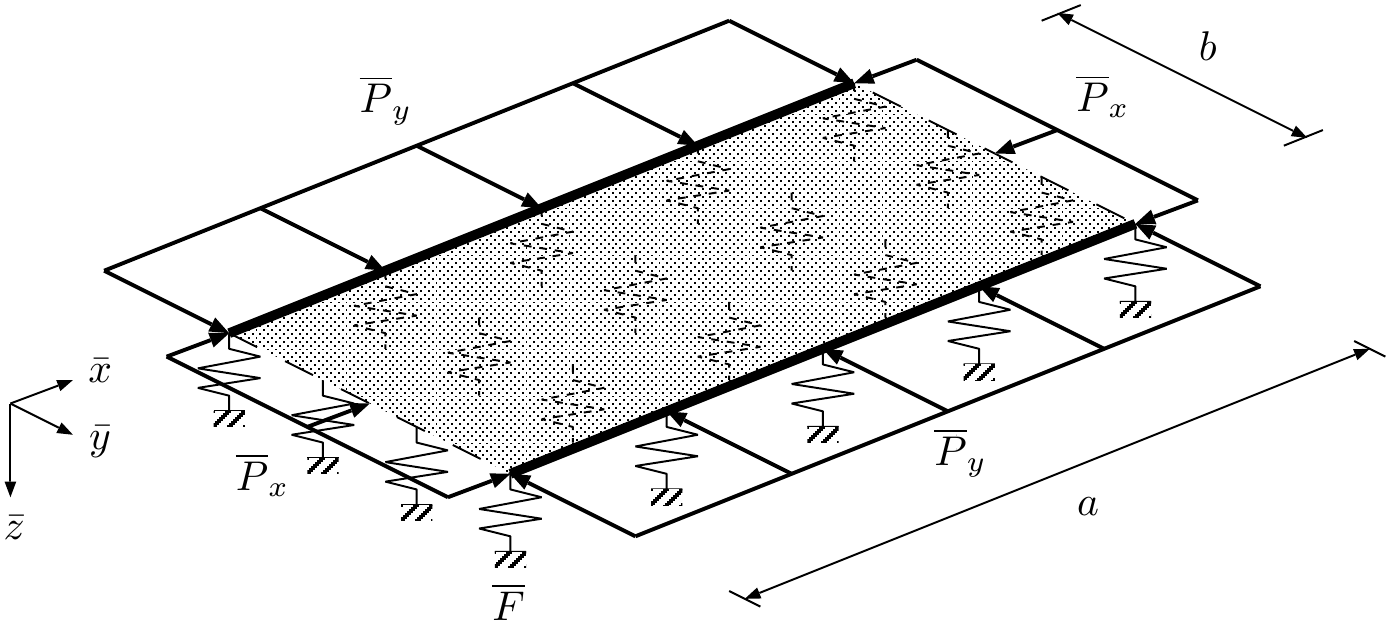}  
  \caption{Schematic representation of a thin rectangular plate
    supported on an elastic foundation.  Fixed boundary conditions are
    applied on the long edges (denoted by the thick lines) while
    Neumann conditions are imposed on the shorter sides (dashed
    lines).  The in-plane loading is $\bar{P}_{x}$ and $\bar{P}_{y}$
    in the $\bar{x}$ and $\bar{y}$ directions respectively.  The
    continuous foundation is represented as a series of elastic
    springs with resistance $\bar{F}$ given by~(\ref{eq:foundation})}
  \label{fig:schematic}
\end{figure}

We consider a thin flat plate made of a homogeneous and isotropic
material of Young's modulus $E$, Poisson's ratio $\nu$ and with
thickness $h$ as shown in Figure~\ref{fig:schematic}.  The co-ordinate
axes are aligned so that the mid-surface of the plate sits in the
$\bar{x}$-$\bar{y}$ plane and the structure is supported in the
$\bar{z}$-direction by an elastic foundation. Buckling is described by
a system derived from the classical von~K\'arm\'an equations
\cite{TimWo} with the addition of terms accounting for the supporting
elastic foundation together with an acceleration term based on the
time $\bar{t}$ \cite{BaZu}. The incorporation of this acceleration
will enable us to later examine the temporal stability of solutions
and then the key equations may be cast as
\begin{subequations}\label{e:FVK}
\begin{align}
  \rho h\bar w_{\bar{t}\bar{t}}
  +D\overline\nabla^{4}\bar{w}
  -h\left(\bar{\phi}_{\bar{y}\bar{y}}\bar{w}_{\bar{x}\bar{x}}
  +\bar{\phi}_{\bar{x}\bar{x}}\bar{w}_{\bar{y}\bar{y}}
  -2\bar{\phi}_{\bar{x}\bar{y}}\bar{w}_{\bar{x}\bar{y}}\right)
  +\bar{P}_{x}\bar{w}_{\bar{x}\bar{x}}+\bar{P}_{y}\bar{w}_{\bar{y}\bar{y}}
  +\bar{F}(\bar{w})&=0,\\
  \overline\nabla^{4}\bar\phi -E\left(\bar{w}_{\bar{x}\bar{y}}^{2}
  -\bar{w}_{\bar{x}\bar{x}}\bar{w}_{\bar{y}\bar{y}}\right)&=0,
\end{align}
\end{subequations}
in which $\overline\nabla^{4}$ denotes the biharmonic operator with
respect to $\bar{x}$ and $\bar{y}$ and all subscripts on variables
(but not parameters) denote partial differentiation with respect to
the appropriate variable.  Moreover $D=Eh^{3}/(12(1-\nu^{2}))$ denotes
the bending stiffness of the plate, $\bar w(\bar x,\bar y,\bar t)$ is
its lateral displacement and $\bar\phi(\bar x,\bar y,\bar t)$ the
corresponding stress function. We note that the terms proportional to
$\bar P_{x}$ and $\bar P_{y}$ account for the in-plane loading in the
$\bar x$ and $\bar y$ directions respectively.

The crucial difference between our model and that for a plate
supported only on its edges is the presence of the foundation
resistance $\bar{F}(\bar w)$ that acts across the plate. Here,
following the methodology used for the corresponding strut problem, we
take $\bar F$ to be of Winkler type \cite{HW} so that it is solely
dependent on lateral deflection $w$ and no shear interaction is
present; then
\begin{equation}
  \bar F(\bar{w})=\bar{k}_{1}\bar{w}-\bar{k}_{2}\bar{w}^{2}
  +\bar{k}_{3}\bar{w}^{3},\label{eq:foundation}
\end{equation}
where the $\bar{k}_j$ are constants and and whose values we discuss
below.

Let us take the plate to be rectangular of dimensions $a\times b$ with
the thickness $h\ll a,b$. Appropriate physical boundary conditions can
then be of various types.  The edges may be unsupported (free),
simply-supported (that prevents deflection but allows rotation) or
fully fixed.  Each one of these conditions supplies two kinematic
conditions on the deflection or its derivatives.  We will study the
case of clamped edges in the $\bar y$-direction and a periodic
displacement along the $\bar x$-direction to simulate a thin layer
bonded to the elastic foundation \cite{CaoW1}.  The cases we will
study will consist of a plate with one of its centre lines on the
$\bar x$-axis with length $a$ and width $b$
\begin{align}
  \bar{w}(\bar x,\pm b/2)=
  \bar{w}_{\bar{y}}(\bar x,\pm b/2)&=0,\nonumber\\
  \bar{w}_{\bar{x}}(0,\bar y)=
  \bar{w}_{\bar{x}}(a,\bar y)=
  \bar{w}_{\bar{x}\bar{x}\bar{x}}(0,\bar y)=
  \bar{w}_{\bar{x}\bar{x}\bar{x}}(a,\bar y)&=0\,,
  \label{eq:BCs}
\end{align}
with similar conditions imposed on $\bar{\phi}$.  Note that the
conditions along the $\bar{x}$-axis resemble the
\emph{symmetric-section} conditions used when seeking localized
solutions to the one-dimensional strut model \cite{HW}.  We will
consider plates of various length in the $\bar{x}$ direction but, if
periodic solutions are sought, then it suffices to use Neumann
boundary conditions as above.

Let us pause here to consider the choice of the nonlinear foundation,
$\bar F$, with a softening characteristic.  Of course linearity is
often an assumption made in order to simplify calculation or to deduce
first-order behaviour.  If we wish to delve deeper into phenomena then
higher-order terms need to be considered and if we write out the
response of the foundation as a Taylor series then it would seem
natural next to incorporate a quadratic term.  However, if we wish to
explore a symmetric response with respect to vertical displacement
then at least a cubic terms needs to be included as well. To decide
the appropriate form of this term it should be remembered that in
civil engineering contexts the foundation medium is frequently some
form of soil.  Most soils are a complex mixture of mineral particles
and varying degrees of water content packed together in an irregular
manner with many voids present.  The interaction of solid granular
particles, water and air is difficult to model but bulk properties can
be derived and are often represented as a linear plus a softening
leading-order nonlinear term with a negative coefficient.  Where other
structural systems such as sandwich panels have been modelled using a
strut- or plate-on-foundation model similar softening nonlinearities
have been introduced into the foundation to good effect
\cite{HdS,HuAhm}.  The rationale behind such an approximation is given
in terms of the bulk behaviour of the central core material which is
often a foam and its voids initially undergo collape when loading but
then may subsequently restabilize due to contact of the solid walls of
the material.  The restabilization is modelled by a higher degree
polynomial term with a positive coefficient.  In material manufacture,
the response of an elastic substrate supporting the material has a
complicated interaction relationship.  To first order, it is often
modelled as linear but as deformations develop, nonlinear responses
are often observed \cite{CaoW1} and can have significant effect on the
overall behaviour of the material.  In both of the above scenarios,
the softening is only the initial deviation from nonlinearity for
there is often some recovery of stiffness (restabilization) when
higher-order kinematic effects are taken into account.  We model this
as the next order (cubic) term having a positive sign.

\subsection{Non-dimensionalization}
In order to analyse the system, it is helpful to scale variables to
remove as many coefficients as possible.  Therefore we set
\begin{equation}
  \bar x=\lambda x,\quad\bar y=\lambda y,\quad
  \bar w=\mu_1w,\quad
  \bar \phi=\mu_2\phi,\quad
  \bar t=\xi t
\end{equation}
where
\begin{align}
  \lambda&=\sqrt[4]{\frac{D}{\bar{k}_1}},\quad
  \mu_1=\sqrt{\frac{D}{Eh}},\quad\mu_2=\frac{D}{h},\quad
  \xi=\sqrt{\frac{\rho h}{\bar{k}_{1}}},\nonumber\\
  P_x&=\frac{\bar{P}_x}{\sqrt{\bar{k}_1D}},\quad
  P_y=\frac{\bar{P}_y}{\sqrt{\bar{k}_1D}},\quad
  k_2=\frac{\bar k_2}{\bar k_1}\sqrt{\frac{D}{Eh}},\quad
  k_3=\frac{\bar k_3D}{\bar k_1Eh}.
\end{align}
This leads to 
\begin{align}
  w_{tt}+\nabla^{4}w-\left(\phi_{yy}w_{xx}+\phi_{xx}w_{yy}-2\phi_{xy}w_{xy}\right)
  +P_{x}w_{xx}+P_{y}w_{yy}
  +F(w)&=0\,,\label{eq:vK1}\\
  \nabla^{4}\phi-\left(w_{xy}^{2}-w_{xx}w_{yy}\right)&=0,\label{eq:vK2}
\end{align}
where $\nabla^4$ is the biharmonic operator with respect to the scaled
variables $x$ and $y$ and $F(w) = w - k_2w^2 + k_3w^3$.  In this
scaled state, we suppose that the plate has length $L$ and width
$\gamma$.

\subsubsection{Loading parameters and some simplifications}
In the general nonlinear plate problem the loading parameters in the
$x$ and $y$ directions, $P_{x}$ and $P_{y}$, should be treated as
independent.  There is no necessity for either to be constant in space
and such variation is likely to be of interest in provoking other
behaviours such as bending about a lateral axis.  We defer this extra
degree of freedom as a refinement to be dealt with in a future study
and instead concentrate solely on the case where both $P_{x}$ and
$P_{y}$ are constant.  That said, we will allow the possibility of
either load being compressive (positive) or tensile (negative).  It is
conceptually easy to envisage different loading in the two directions:
the simplest of course being uniaxial compression.  Furthermore if the
edges are prevented from moving in the plane by a pin or clamped
support then tensile stresses can be built up as the plate deforms in
the $z$ direction as moderate to large buckling deformation increases.
As outlined in Yang \emph{et al.}  \cite{Yang15}, one-dimensional
wrinkles can be induced in nanomaterials by pre-stretching the
substrate and then depositing the material of interest on it. When the
tensile force is released this results in a compression acting in one
direction. Forces can then be applied in an orthogonal direction to
induce further changes to the material.

In this study we will limit investigation to the case when, say,
$P_{y}$ is held constant and $P_{x}$ is parametrically varied.
Physically this is equivalent to preloading the plate in one direction
(not necessarily up to the buckling capacity in that direction) and
then applying another load in the orthogonal direction.

\subsection{Total potential energy and Euler--Lagrange equations}
We shall use the energy of the system as a measure of the various
buckled states. The total potential energy associated with the
non-dimensional form of the von~K\'arm\'an equations in a slightly
compacted form is \cite{EvHu99}
\begin{align}
  V(w,\phi)=&\frac{1}{2}\int\int\left[(\nabla^{2}w)^{2}-2(1-\nu)\left(
      w_{xx}w_{yy}-w_{xy}^{2}\right)\right]\D x\,\D y\nonumber\\
  &+\frac{1}{2}\int\int\left[-(\nabla^{2}\phi)^{2}-2(1+\nu)\left(
      \phi_{xx}\phi_{yy}-\phi_{xy}^{2}\right)\right]\D x\,\D y\nonumber\\
  &-\int\int\phi\left[-w_{xy}^{2}+w_{xx}w_{yy}\right]\D x\,\D y\nonumber\\
  &-\frac{1}{2}\int\int\left[P_{x}w_{x}^{2}
    +P_{y}w_{y}^{2}\right]\D x\,\D y\nonumber\\
  &+\int\int\left[\frac{1}{2}w^{2}-\frac{1}{3}k_{2}w^{3}
    +\frac{1}{4}k_{3}w^{4}\right]\D
  x\,\D y.\label{eq:V}
\end{align}
The first three lines of integrals represents the energy stored in the
bending of the plate and includes nonlinear contributions arising from
moderate to large deformations.  The fourth line denotes the work
done by the two loading parameters $P_{x}$ and $P_{y}$ and we identify
the end-shortening of the plate in the $x$ and $y$ directions to be
\begin{equation}
  \Delta_{x}=\frac{1}{2}\int\int w_{x}^{2}\,\D x\,\D y
  \quad\text{and}\quad
  \Delta_{y}=\frac{1}{2}\int\int w_{y}^{2}\,\D x\,\D y,
  \label{eq:endsh}
\end{equation}
respectively.  The final line of integrals in (\ref{eq:V}) gives the
energy stored in the elastic foundation.

Of course, the governing equations are obtainable from $V$ by
obtaining the associated Euler--Lagrange equations of the system but
the energy itself plays an important role in the preferential
formation of certain buckle patterns.

\subsection{Linearization}
To understand the critical behaviour of the system it is necessary to
linearize~(\ref{eq:vK1})-(\ref{eq:vK2}) about the trivial state. We
will first look at the critical instability of an infinite plate so
that boundary conditions can be safely ignored. In order to simplify
the subsequent notation let us put $P_x=P$ and $P_y=\alpha P$ for
where the parameter $\alpha$ can theoretically be any real
value. However, without loss of generality, if we consider a plate
that is longer in the $x$-direction than in the $y$ then buckling will
be dictated by $P_x$ and so we can legitimately restrict our interest
to $|\alpha|\leq1$. If we then seek Fourier mode solutions
\begin{equation}
  w(x,y)=A\E^{\Lambda x}\E^{\im\pi y},
\end{equation}
of the linearization of
~(\ref{eq:vK1})-(\ref{eq:vK2}) about the trivial state, this gives
\begin{equation}
  (\Lambda-\pi)^2(\Lambda+\pi)^2+(\Lambda^2-\alpha\pi^{2})P+1=0,
\end{equation}
with solutions
\begin{equation}
  \Lambda=\pm\frac{\sqrt{\pm\sqrt{S}
      -P+2\pi^2}}{\sqrt{2}}\qquad S\equiv P^2-4\pi^2P(1-\alpha)-4.
\end{equation}
A bifurcation occurs when $S$ becomes negative and this then
corresponds to $x$-spatially periodic modes.  When $P<\Pc$ the
eigenvalues are complex-valued with the critical values $\Pc=2$ when
$\alpha=1$ and $\Pc=2\pi^{2}+\sqrt{\pi^{4}+1}$ when $\alpha=0$; in
contrast when $P>\Pc$ two of the eigenvalues are complex but the other
two real.  Of course we should also linearize equation~(\ref{eq:vK2})
but as the result is independent of $P$ it has no bearing on the
bifurcation behaviour of the system with respect to load.

\section{A Galerkin analysis}
\label{sec:galerkin}
To obtain numerical solutions and then investigate their evolution
under varying parameters it is necessary to derive an initial
approximate two-dimensional solution to the system.  Once a successful
solution is obtained, we can employ numerical continuation to change
$P_{y}$ to the desired value.

We utilize a straightforward Galerkin procedure \cite{WaHiHu} for this
purpose and start with a purely periodic response of the
form
\begin{equation}
  w=A\psi(x,y)\label{eq:w}
\end{equation}
where $A$ is a constant amplitude and $\psi$ is an appropriate shape
function.  The initial stress function, $\phi$, can be approximated
either as a multiple of $w$ or as zero if we are seeking to start from
close to the bifurcation from the trivial state.  We set the length of
the plate in the $x$ direction to be $L=\pi$ and impose Neumann
boundary conditions in that direction.  If the width $\gamma=2$, then
for clamped boundary conditions in the $y$-direction, at $y=\pm1$, we
need to identify an appropriate set of functions that satisfy
$w(x,\pm1)=w_{y}(x,\pm1)=0$.  In the $x$-direction we assume Neumann
boundary conditions, i.e., $w_{x}(0,y)=w_{x}(\pi,y)=0$ and,
furthermore, if $w(0,y)=w(\pi,y)$ then we have periodic boundaries in
that direction.  A suitable family of even functions obeying these
boundary conditions is given by
\begin{equation}
  \psi(x,y)=\cos m x((-1)^{n+1}+\cos n\pi y)\label{eq:even}
\end{equation}
where $m$ and $n$ are integers which characterize the number of half
waves occuring in the $x$ and $y$ directions, respectively and a
periodic boundary condition in the $x$-direction requires $m$ to be
even.  A counterpart odd set of functions in the $y$ direction are
given by
\begin{equation}
  \psi(x,y)=\cos m\pi x\left(\sin n\pi y
    +\frac{n}{n+1}\sin(n+1)\pi y\right)\label{eq:odd}
\end{equation}
and in order to satisfy periodic boundary conditions in the $x$
direction, $m$ should be even.  If we denote the left-hand side of the
first equation (\ref{eq:vK1}) by $Q(x,y)$ and feed in the assumed form
for $w$ in (\ref{eq:w}) then the Galerkin formulation tells us that
\begin{equation}
  \int_{y=-1}^{y=1}\int_{x=0}^{x=\pi}Q(x,y)\psi(x,y)\,\D x\,\D y=0\label{eq:gal}
\end{equation}
which leads to a cubic equation in $A$ with a non-trivial even solution 
\begin{equation}
  A=\pm\frac{4}{\sqrt{105k_{3}}}\sqrt{3m^{2}P_{x}+\pi^{2}n^{2}P_{y}
    -(3m^{4}+2\pi^{2}m^{2}n^{2}+\pi^{4}n^{4}+3)}.
\end{equation}
Note the presence of the cubic coefficient $k_{3}$ but the absence of
the quadratic coefficient $k_{2}$.  This feature is a consequence of
taking a single term for the shape of $w$ and leads to a symmetric
from for $A$.  For greater accuracy we could use more terms but since
the motivation for this Galerkin method is simply to supply some
initial functions that can be used to generate full numerical
solutions close to the trivial one, there is little to be gained by
deriving a more accurate initial state.  For real values of $A$ the
term inside the square root must be greater than zero and this gives
us a critical value for $P_{x}=P_{x}^{C}$, say, if $P_{y}$ is held
constant.  %Thus we
%obtain the expression
%\begin{equation}
%  P_{x}^{C}\geq\frac{3m^{4}+2\pi^{2}m^{2}n^{2}+\pi^{4}n^{4}+3
%   -\pi^{2}n^{2}P_{y}}{3m^{2}}.\label{eq:galPx}
%\end{equation}
%Each of these approximations can be used as initial solutions for
%numerical solutions (see \S\ref{sec:num}).  Of course, more accuracy
%would be achieved by using more than a single degree of freedom but
%our aim here has been to show that two-dimensional periodic forms
%arise in the post buckling of the plate resting on a foundation.

% {\bf D: I think it would be interesting to use the same ansatz but to
%   plug it into the energy functional and minimise with respect to the
%   amplitudes of the ansatz for $w$ and $\phi$. We can then compare
%   with the Galerkin analysis and also compute stability.}

% {\bf D:Comment on stability?}

\section{Transverse instability of quasi-1D states}
\label{sec:stab}
We next examine the possibility of a transverse instability of
quasi-1D states (by which we mean one-dimensional states that are
trivially extended to two-spatial dimensions). We will subsequently
find that the transverse instability of quasi-1D states play a pivotal
role for the emergence of non-trivial 2D states.

We start by seeking solutions that vary only in the transverse
direction $y$.  If $W(y,t)$ is a one-dimensional solution of
(\ref{eq:vK1}) and if $\Phi(y)$ satisfies (\ref{eq:vK2}) then it is
immediately apparent that
\begin{align}
 W_{tt} +  W_{yyyy}+P_{y}W_{yy}+f(W)&=0,\label{eq:W-ode}\\
  \D_{y}^{4}\Phi&=0.\label{eq:phi-ode}
\end{align}
In such state, as there is no variation in $x$, the end-shortening
component $\Delta_{x}=0$ and the stress function $\Phi\equiv 0$
everywhere. It is well known that stationary solutions
of~(\ref{eq:W-ode}), $W(y,t)=W(y)$, exist; see, for instance,
see~\cite{PelTr}. The linear stability properties of the stationary
$W(y)$ with respect to perturbations in $y$ can be inferred directly
from results of the ubiquitous SH equation (given by (\ref{eq:W-ode})
save that the $W_{tt}$ is replaced by $W_t$). In particular, if a
periodic orbit is linearly stable in the standard SH equation, then it
is neutrally stable in~(\ref{eq:W-ode}). Hence, there do exist
periodic orbits that are linearly stable with respect to perturbations
in the $y$-direction.

We next consider a small transverse perturbation in the $x$-direction
by letting $w(x,y,t)=W(y)+\hat{w}(x,y,t)$ and
$\phi(x,y,t)=\hat{\phi}(x,y,t)$, where $\hat{w}$ and $\hat{\phi}$ are
both small. Substituting these forms in (\ref{eq:vK1}) and
(\ref{eq:vK2}) and linearizing we find
\begin{align}
  \hat{w}_{tt}+%\hat{w}_{xxxx}+2\hat{w}_{xxyy}+\hat{w}_{yyyy}
  \nabla^{4}\hat{w}
  -(\hat{\phi}_{xx}\,W_{yy})
  +P_{x}\hat{w}_{xx}+P_{y}\hat{w}_{yy}+f'(W)\hat{w}&=0,\\
  %\hat{\phi}_{xxxx}+2\hat{\phi}_{xxyy}+\hat{\phi}_{yyyy}
  \nabla^{4}\hat{\phi}
  +\hat{w}_{xx}W_{yy}&=0.
\end{align}
We look for periodic perturbations in the $x$-direction so that
\begin{equation}
  \Bigl\{\hat w, \hat \phi\Bigr\}=\exp(\eta t
  +\im\beta x)\Bigl\{\hat{v}(y),\hat{\psi}(y)\Bigr\}+\textrm{c.c.},
\end{equation}
where c.c. denotes the complex conjugate, it follows that
$\hat{v}$ satisfies the ODE eigenvalue problem
\begin{align}
  \eta^{2}\hat{v}+\cL_{1}(\hat{v})+\beta^{2}\hat{\psi}W_{yy}
  +\beta^{4}\hat{v}-2\beta^{2}\hat{v}_{yy}
  -\beta^{2}P_{x}\hat{v}&=0\label{eq:evp}\\
  \cL_{2}(\hat{\psi})-2\beta^{2}\hat{\psi}_{yy}+\beta^{4}\hat{\psi}
  -\beta^{2}\hat{v}W_{yy}&=0
\end{align}
in which the operators are
\begin{equation}
  \cL_{1}(\hat{v}):=\hat{v}_{yyyy}+P_{y}\hat{v}_{yy}+f'(W)\hat{v},\quad
  \cL_{2}(\hat{\psi}):=\hat{\psi}_{yyyy}.
\end{equation}
%Since we are on the infinite plate and due to translational symmetry,
%we know that $\cL_{1}(W_{y})=0$.

We solve the spectral problem (\ref{eq:evp}) for
$\eta=\eta(\beta)$.  For small $\beta$ we write
\begin{equation}
  (\eta,\hat{v},\hat{\psi})=(0,\hat{v}_{0}(y),0)
  +\beta(\eta_{1},\hat{v}_{1}(y),\hat{\psi}_{1}(y))
  +\beta^{2}(\eta_{2},\hat{v}_{2}(y),\hat{\psi}_{2}(y))+\cdots
\end{equation}
and at the zeroth and first orders it follows that $\hat{v}_0=W_y$ and
$\cL_1(\hat{v}_1)=\cL_2(\hat{\psi}_1)=0$ so that $\hat{v}_1\equiv0$ and
$\hat{\psi}_1\equiv0$.

At  $\mathcal{O}(\beta^2)$, we have
\begin{equation}\label{e:beta2}
  \cL_{1}(\hat{v}_{2})=2\hat{v}_{0,yy}+P_{x}\hat{v}_{0}
  -\eta_{1}^{2}\hat{v}_{0},\qquad
  \cL_{2}(\hat{\psi}_{2})=\hat{v}_{0}\,W_{yy}.
\end{equation}
A standard application of the method of adjoints leads to the
conclusion that this system only admits an acceptable solution if
\begin{equation}\label{eq:solveability}
  \eta_{1}^{2}N_{1}^{2}+2N_{2}^{2}-P_{x}N_{1}^{2}=0,
\end{equation}
where $N_{1}\equiv\int W_{y}^{2}\,\D y$ and
$N_{2}\equiv\int W_{yy}^{2}\,\D y$.  Therefore the stability boundary
is given by $\eta=\eta_{1}\beta+\mathcal{O}(\beta^2)$ with
\begin{equation}
  \sgn\eta_{1}^{2}=\sgn(N_{1}^{2}P_{x}-2N_{2}^{2}).\label{eq:eta1}
\end{equation}
This criterion for transverse instability replicates that for the
standard SH equation~\cite{hoyle2006}. In particular, it is known that
1D periodic orbits minimise their energy with respect to variation in
their wavelength.

\begin{figure}[h]
  \centering
  \includegraphics[width=0.8\textwidth]{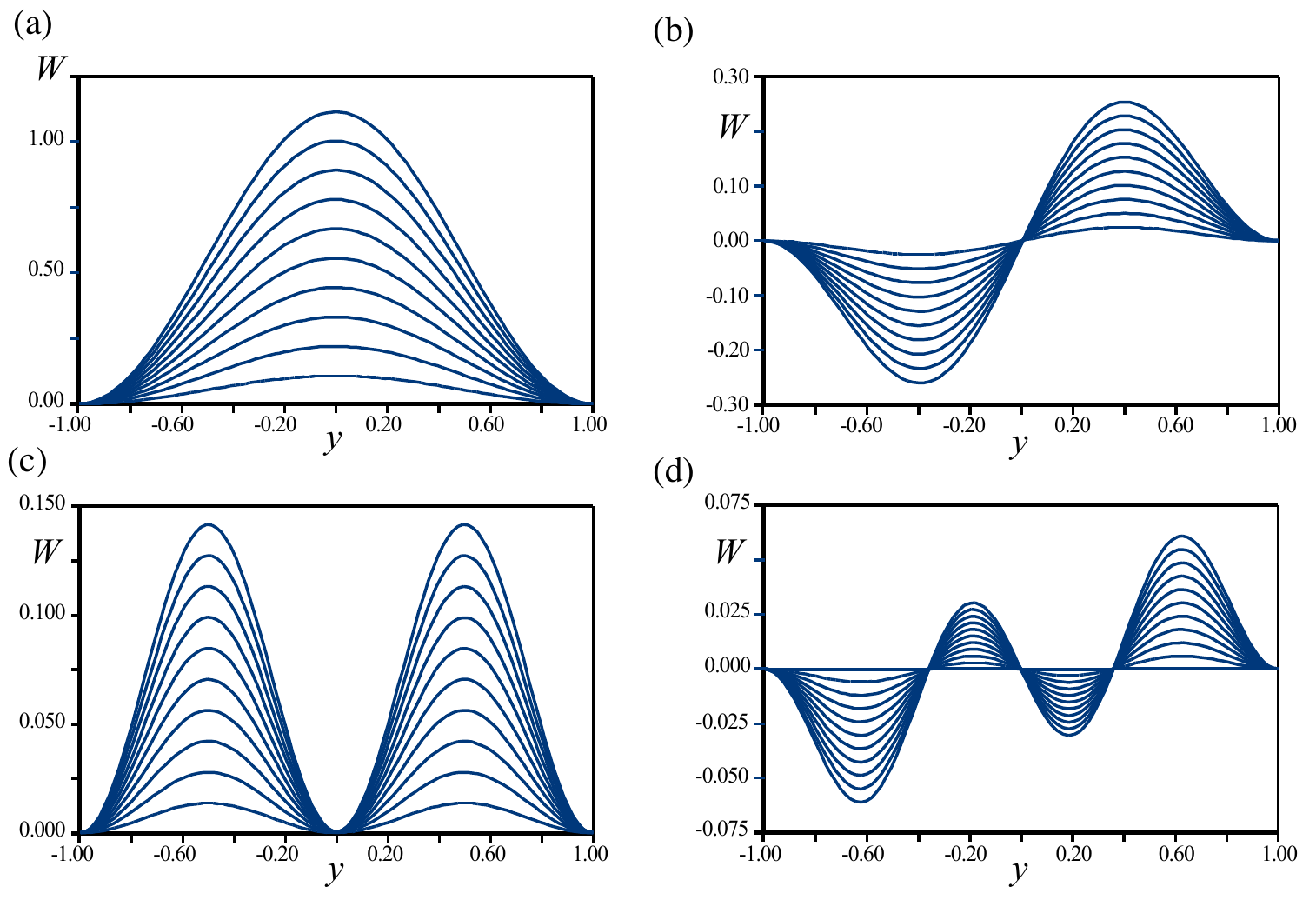}
  \caption{Steady solutions of equation (\ref{eq:W-ode}) with
    $f(W)=W-2W^{2}+W^{3}/2$, a normalized length $L=2$ and various values of
    $P_{y}$. (a)~first-mode, $\Pc=10.17$, (b)~second-mode,
    $\Pc=20.27$, (c)~third-mode, $\Pc=39.55$ and (d)~fourth-mode,
    $\Pc=59.71$.}
  \label{fig:modes}
\end{figure}

We now have a procedure to predict the dynamical stability of
patterns, at least when they start as functions of $y$ alone.  We
therefore need to obtain solutions, $W(y)$, of the ODE
(\ref{eq:W-ode}), subject to appropriate boundary conditions (clamped
in this case).  Solutions can readily be obtained using
\textsc{Auto}~\cite{Automan07} and the quantities $N_1$ and $N_2$ used
to evaluate the transverse instability
criterion~(\ref{eq:eta1}). % Using \textsc{Auto}, it is possible to
%obtain $N_{1}^{2}$ and $N_{2}^{2}$ directly.  
We compute the first four modes of~(\ref{eq:W-ode}) on a finite domain
in $y$ satisfying the boundary conditions~(\ref{eq:BCs}) and the
results are shown in
Figure~\ref{fig:modes}. %These modes can be extended periodically on
%the infinite line where they can then be arbitrarily shifted.

%The above results indicate that if $\eta_{1}$ is imaginary then 5the amplitude of the deflection $\hat{w}$ does not grow
%n time.  Note, since the system is conservative (no dissipation
%term---or $w_{t}$ is present in (\ref{eq:vK1})) meaning that any real
%part of $\eta_{1}$ would cause instability (growth) whether in
%forward or backward time.

\begin{figure}[h]
  \centering
  \includegraphics[width=0.5\textwidth]{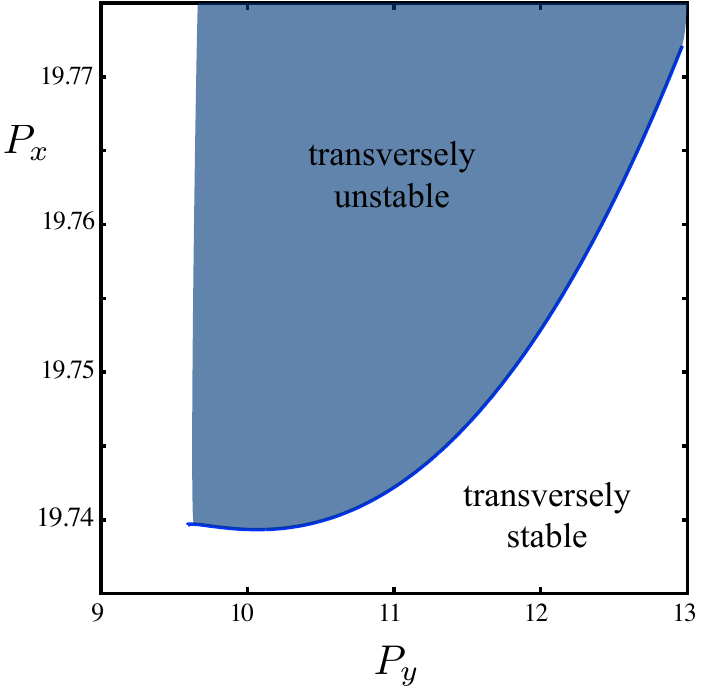}
  \caption{Two-parameter plot of the transverse instability region for
    the first mode in Figure~\ref{fig:modes} for $k_2=2$ and
    $k_3=0.5$.  The line shown marks the change in sign of
    $\eta_{1}^{2}$ (\ref{eq:eta1}) from negative (unshaded region) to
    positive (shaded region); the former are stable and the latter
    unstable.  Note that the first mode cannot exist once
    $P_y\lessapprox9.7$.\label{fig:two_param_trans}}
\end{figure}

We show in Figure~\ref{fig:two_param_trans} a two-parameter plot of
the transverse (in)stability region when $k_2=2$ and $k_3=0.5$.  In
this figure the values of $N_{1}$ and $N_{2}$ (see
(\ref{eq:solveability})) have been calculated numerically for a
one-dimentional pattern and the boundary where the value of
$\eta_{1}^{2}$ becomes positive (\ref{eq:eta1}) is shown.  It
indicates the value of $P_{x}$ above which the solutions become
unstable in time.  We see that the location of the lower boundary for
$P_y$ is almost independent of $P_x$; indeed when $P_y\lessapprox9.7$,
the first-mode no longer exists and bifurcates from the flat state.
We infer from this result that the striped (one-dimensional) solution
remains stable until $P_{x}$ is pushed beyond some ctitical value,
after which it becomes unstable and the pattern evolves to some other
state.

A word of caution is appropriate here. Our analysis described here is
strictly only valid for a plate of infinite extent in the
$x$-direction. Thus it is unlikely to provide a complete picture of
the stability of plates of finite length for which the choice of
boundary conditions ought to be relevant. That said, the analysis
should provide some clue as to the types of behaviours that may be
seen when a striped solution of a finite plate (with deflection
independent of $x$) encounters a perturbation which varies in the
transverse planar direction.

\section{Numerical solutions}
\label{sec:num}
The above approximations allow us to speculate on the form of some of
the possibilities of solution of the plate-on-foundation model.
However, given the experience gained with related problems, such as
the (1D) strut model \cite{WaBa}, it is highly probable that there is
an intricate solution structure which, initially at least, can best be
explored numerically. This was done by discretising the boundary value
problem using a standard fourth order scheme in the $x$-direction and
then imposing Neumann boundary conditions using ghost
points~\cite{smith2003}. A pseudo-spectral Chebyshev discretisation
was employed in the $y$-direction with the clamped boundary conditions
enforced by choosing the polynomial interpolant, $p(y)$, of a function
in $y$ to be $p(y)=(1-y^2)q(y)$ where $q(y)$ is a polynomial with
$q(\pm1)=0$; see Trefethen \cite[Chapter 14]{trefethen2000}. The
two-dimensional differentation operators were constructed using
Kronecker products as described in~\cite{trefethen2000} and the
boundary value problem solved using a Newton trust region based
algorithm~\cite{coleman1996}. We also embedded the boundary value
problem into a secant continuation formulation to facilitate path
following of the solutions in parameter space~\cite{krauskopf2007}.

The scheme was implemented in \textsc{Matlab} (version 2015a) with
typical discretizations in $x$ of $N_x=41$ mesh points on
$x\in[0,\pi]$ and $N_y=21$ Chebyshev collocation points in 
$y$.

\subsection{Periodic solutions}
As mentioned earlier, it is frequently the case that for plate
problems focus is concentrated on periodic behaviour particularly in
the macroscopic example of engineering structures.  In such instances
the wavelength of the buckle pattern is of the same order of magnitude
as the overall dimensions of the plates so only a relatively small
number of waves are expected to span the whole domain. The aspect
ratio is also important; a plate that is very much thinner in one
dimension than the other might conceivably behave somewhat like a
strut \cite{WHW} but wider plates may well exhibit very different
behaviour.  We take some tentative numerical steps towards
investigating this variation in \S\ref{sec:num}\ref{sec:aspect}.

\begin{figure}[h]
  \centering
  \includegraphics[width=\textwidth]{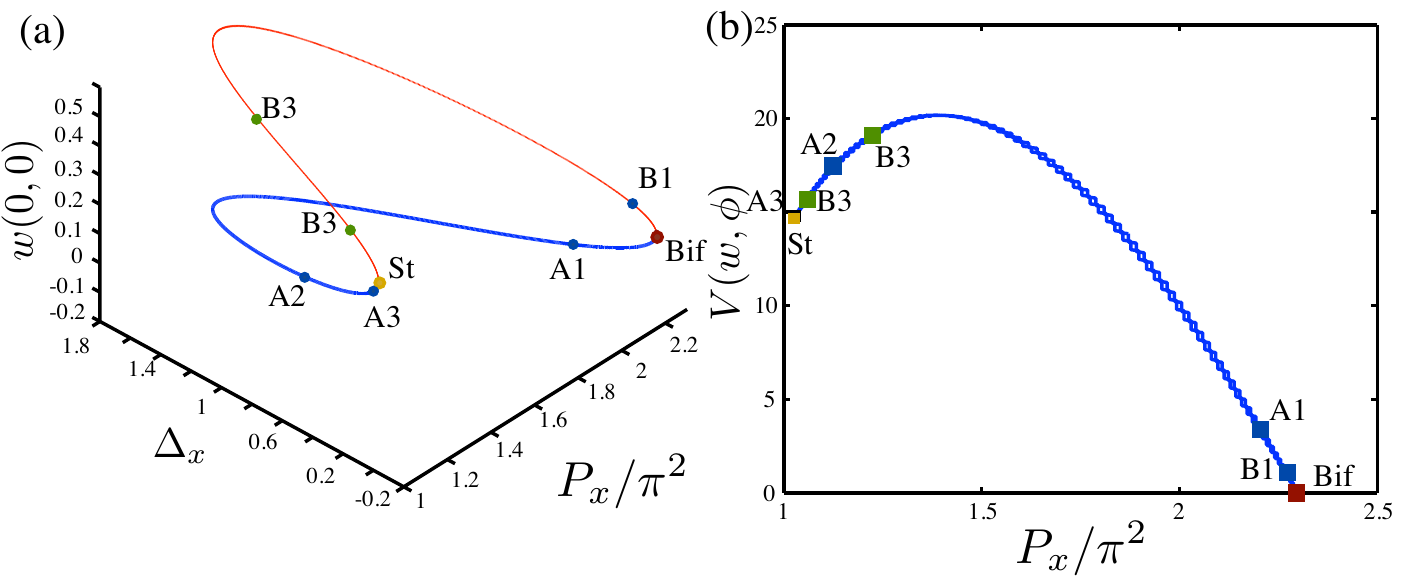}
  \caption{(a)~Bifurcation diagram and (b)~potential energy plot of
    periodic solutions of the von~K\'arm\'an plate-on-foundation model
    under in-plane forces with $k_{2}=2$ and $k_{3}=0.5$.  The plate
    has length $L=\pi$ and the transverse load, $P_{y}$, is held
    constant at 2.  The two branches are denoted~A in blue (thick
    line), and~B in red (thin line).  (b)~In $P_{x}$-$V$ space the two
    branches are congruent and solutions corresponding to individual
    points identified on each branch are shown in
    Figures~\ref{fig:num1} and \ref{fig:num7}.}
\label{fig:bif1}
\end{figure}

The earlier Galerkin analysis suggested that solutions which are
periodic both in $x$ and $y$ can bifurcate from the trivial flat state
and they do so in pairs.  However we now appreciate a limitation of
the simple-minded Galerkin approximation used above because it is
evident that, since the foundation is asymmetric with respect to
deflection the pair of modes that evolve are not mirror images of each
other.  In order to model this asymmetric behaviour, more than a
single mode would be required to break the implicit symmetry
\cite{HBT}.  Nevertheless, the single-mode approach gives us a firm
connection between the one- and two-dimensional studies.  A typical
example of the periodic post-buckling regime is depicted in
Figure~\ref{fig:bif1}.  Here the two branches emerging from the flat
state are labelled $A$ and $B$ and shown in different colours.  In
$P$-$\Delta_{x}$ space both branches overlap exactly and it is
therefore more informative to present the plot with one extra
dimension (here, deflection at the origin) to prise the two branches
apart and show that they really are distinct.  The buckled shape
corresponds to $m=3$, $n=1$ as defined in (\ref{eq:even}).  With
$P_{y}=2$, the prediction from the 1D approach in \S\ref{sec:stab} is
that at bifurcation $P_{x}\approx17.03$ (see~(\ref{eq:eta1})) while
the numerical solution of the full equation reveals that the
bifurcation occurs at around $P_{x}=22.7$.  Each point on the curve is
a solution of the static equilibrium equations although the energy $V$
does vary considerably.  The value of the energy can be loosely
thought of as an informal relative measure of how distorted the
deflected surface is, starting at zero at the bifurcation point (Bif)
and dropping off as the striped solution is approached (St) (although
as the striped solution is obviously not flat, the value of the energy
is not zero).

\begin{figure}[h]
  \centering
  \includegraphics[width=\textwidth]{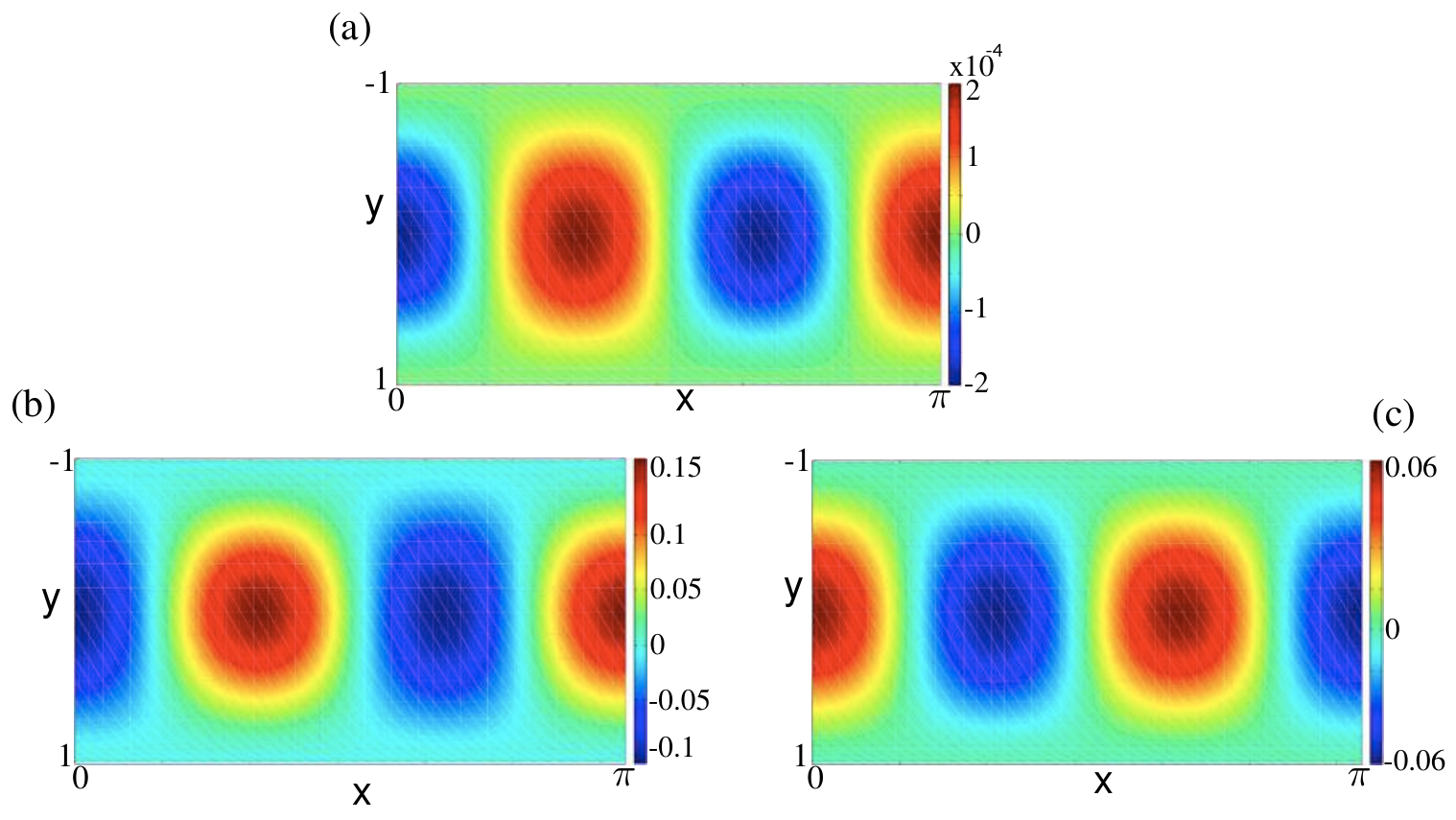}
  \caption{Contours of $w(x,y)$ predicted by numerical solution of the
    von~K\'arm\'an plate-on-foundation equations under in-plane
    forces.  Domain of plate is $x\in[0,\pi]$ and
    $y\in[-1,1]$. (a)~Solution close to bifurcation point, denoted as
    Bif in Figure~\ref{fig:bif1}, $P_{x}=22.695$. % Px=2.2995
    The indicated deformation, $w$, is in the range
    $\pm2\times10^{-4}$. (b)~Solution corresponding to Point~$A_1$ in
    Figure~\ref{fig:bif1}, $P_{x}=21.761$, % Px=2.2048
    $w\in[-0.11,0.17]$. (c)~Solution corresponding to Point~$B_1$ in
    Figure~\ref{fig:bif1}, $P_{x}=22.434$, % Px = 2.273
    $w\in[-0.085,0.082]$.  These solutions correspond to the even case
    with $m=3$, $n=1$ in \S\ref{sec:galerkin}.  }
  \label{fig:num1}
\end{figure}

\begin{figure}[htbp]
  \centering
  \includegraphics[width=\textwidth]{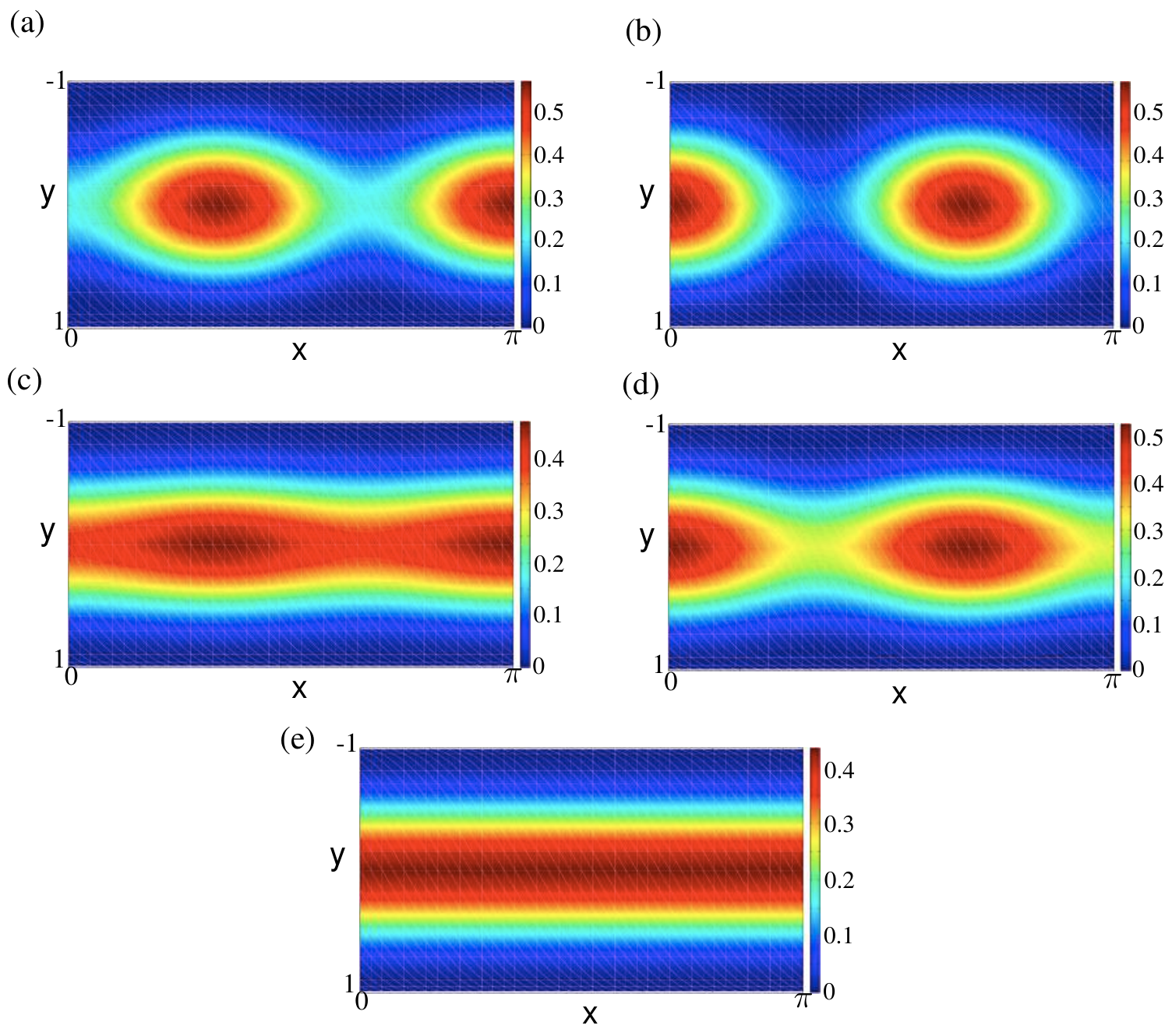}
  \caption{Contours of $w(x,y)$ from numerical solutions using the
    equations and geometry of Figure~\ref{fig:num1}. The five plots
    correspond to (a)~Point~$A_2$ in Figure~\ref{fig:bif1} with
    $P_{x}=11.126$ and $w\in[0,0.57]$. Equivalent data for the
    remaining four plots are (b)~$B_2$, $P_{x}=12.070$ and
    $w\in[-0.01,0.58]$; (c)~$A_3$, $P_{x}=10.153$ and $w\in[0,0.47]$;
    (d)~$B_3$, $P_{x}=10.465$ and $w\in[0,0.55]$; (e)~limit point St,
    $P_{x}=10.110$ and $w\in[0,0.43]$. }
\label{fig:num7}
\end{figure}

Some solutions at various values of $P_x$ are shown in
Figures~\ref{fig:num1} and \ref{fig:num7}.  In
Figure~\ref{fig:num1}(a) we see the buckle pattern, which emerges from
the flat state $w=0$.  Of course, this solution relates to just one
side of the bifurcation point as a second pattern, which is not quite
the mirror image of the first, evolves on the other branch.
Figures~\ref{fig:num1}(b) and (c) depict solutions further into the
post-buckling regime and it is much easier to see that these solutions
are asymmetric about the flat state owing to the bias in the
foundation introduced by the quadratic term.

Solutions on each branch continue to maintain the same structure as
$P_{x}$ is varied except that the amplitudes of both the peaks and
troughs increase steadily until the branches near the striped solution
St which is shown in Figure~\ref{fig:num7}. This last state
corresponds to the situation when the two branches eventually coalesce
thus forming a closed bifurcation curve (isola) on which the solution
appears to be trapped.

\subsubsection{Striped solution}
It is evident that a special example of a periodic solution is that
discussed in \S\ref{sec:stab} where the deflection is purely a
function of $y$.  The solution has been shown to be stable to a
two-dimensional perturbation only if the axial load $P_{x}$ lies
within a certain range. Our computations above indicate that the
solution represents a transition from solutions of one parity to
another.  The numerical results also confirm that $\Delta_{x}$ is zero
both at the bifurcation from the trivial state \emph{and} at the
striped state (see Figure~\ref{fig:bif1}(a)).  A very important
distinction between these two limiting configurations is that the
potential energy is quite different--- see Figure~\ref{fig:bif1}(b).
The upshot is that the two states are not interchangeable and a
straightforward transition from one to the other is impossible.

\begin{figure}
  \centering
  \includegraphics[width=\textwidth]{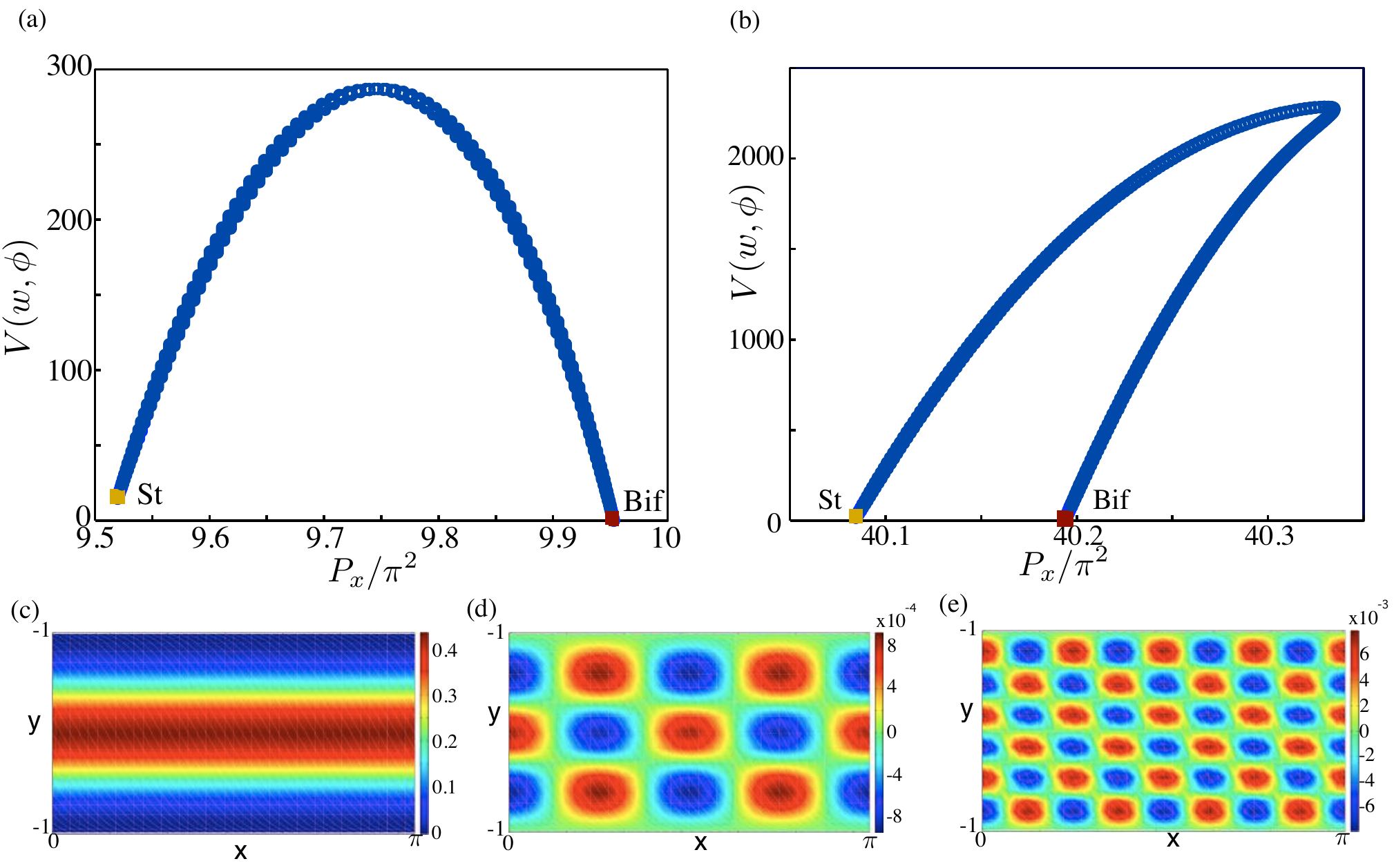}
  \caption{Bifurcation diagrams of two periodic solutions of the
    von~K\'arm\'an plate model under in-plane forces with $k_{2}=2$
    and $k_{3}=0.5$ (a)~even solution with $m=4$, $n=2$ and (b)~odd
    solution with $m=8$, $n=3$ (see \S\ref{sec:galerkin}).  The
    transverse load is $P_{y}=2$.  (c)~Numerical solution at
    coalescent limit points St for both (a) and (b), with
    $w\in[0,0.43]$. (d)~Numerical solution of deformed plate near
    bifurcation point Bif in~(a) with
    $w\in[-9\times10^{-4},9\times10^{-4}]$.  (e)~Numerical solution
    near bifurcation point Bif in~(b) with
    $w\in[-8\times10^{-3},8\times10^{-3}]$.}
\label{fig:bif3}
\end{figure}

For the same value of $P_{y}$ $(=2)$, two further families of
solutions can be found that both bifurcate from the fundamental state
but which do so at different values of $P_{x}$ (see
Figures~\ref{fig:bif3}).  The emerging solutions
are distinct, having a different number of waves in the $y$ direction
but, as these solutions evolve, they converge to the same striped
solution presented earlier.  Though kinematically similar, it should
be remembered that $P_{x}$ is different in the two instances although
the final striped solution has potential energy $V\approx14.649$
(equation~(\ref{eq:V})).  The potential energy of the striped solution
is lower than other equilibrium states in the vicinity which satisfies
the axiom for static stability \cite{TH73} and numerically agrees with
our analysis of dynamical stability in \S\ref{sec:stab} that such
solutions should be statically stable.

% \begin{figure}
% \includegraphics[width=\textwidth]{figure6}
% \caption{(a)~Bifurcation diagram and potential energy plot of periodic
%   solutions of the von~K\'arm\'an plate model under in-plane forces
%   with $k_{2}=2$ and $k_{3}=0.5$ (even solution with $m=4$, $n=2$ in
%   \S\ref{sec:galerkin}).  The transverse load $P_{y}=2$.
%   (b)~Numerical solution of deformed plate near bifurcation point with
%   deformation in the range $\pm9\times10^{-4}$.  (c)~Solution at
%   coalescent limit point.  Deformation in the range $(0,0.43)$}
% \label{fig:bif2}
% \end{figure}

% \begin{figure}
% \includegraphics[width=\textwidth]{figure7}
% \caption{Bifurcation diagram and potential energy plot of periodic
%   solutions of the von~K\'arm\'an plate model under in-plane forces
%   with $k_{2}=2$ and $k_{3}=0.5$ (odd solution with $m=8$, $n=3$ in
%   \S\ref{sec:galerkin}).  The transverse load, $P_{y}=2$.
%   (b)~Numerical solution near bifurcation point with deformation in
%   the range $\pm8\times10^{-3}$. (c)~Numerical solution at coalescent
%   limit point. Deformation in the range (0,0.43).}
% \label{fig:bif3}
% \end{figure}

\begin{figure}
  \centering
  \includegraphics[width=\textwidth]{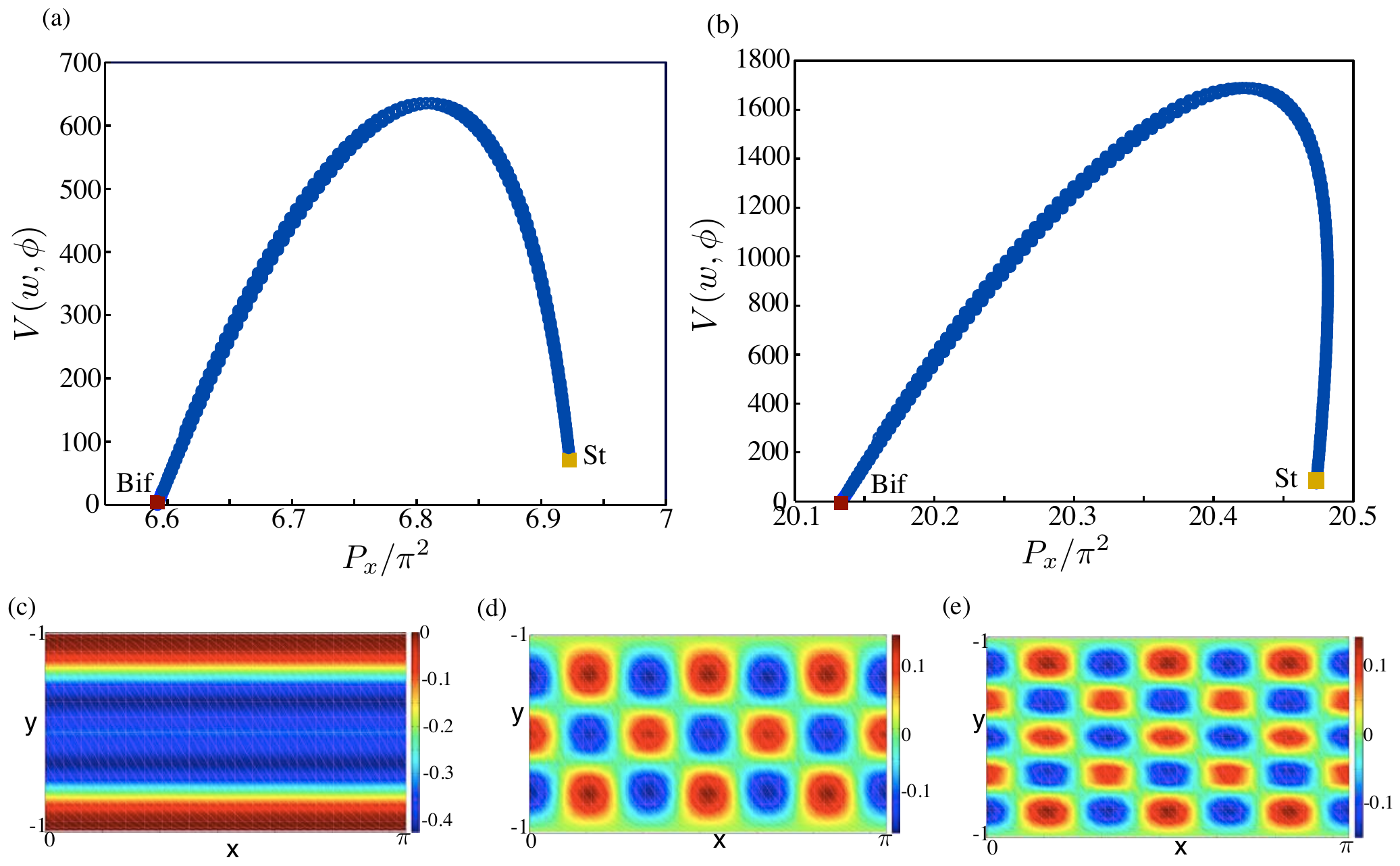}
  \caption{Two bifurcation diagrams of periodic solutions of the
    von~K\'arm\'an plate model under in-plane forces with $k_{2}=2$
    and $k_{3}=0.5$. (a)~An even solution with $m=6$, $n=2$ as defined
    in \S\ref{sec:galerkin}.  (b) An~even solution with $m=6$, $n=3$.
    In both cases, the transverse load, $P_{y}$, is held constant at
    7. (c)~Solution at coalescent limit points shown for both (a)
    and (b) as St with $w\in[-0.42,0]$.  (d)~Numerical solution near
    bifurcation point Bif in~(a) ($w\in[-0.17, 0.2]$). (e)~Numerical
    solution near bifurcation point Bif in~(b) ($w\in[-0.2,0.2]$).}
\label{fig:bif4}
\end{figure}

Two further bifurcation diagrams are presented for a different value
of transverse load: $P_{y}=7$ (see Figure~\ref{fig:bif4}).  It can be
seen once more that although the solutions emergent from the trivial
state have separate forms, they again coalesce onto the same striped
solution.  In this case, $V\approx73.821$.

% \Begin{figure}
% \includegraphics[width=\textwidth]{figure8}
% \caption{Bifurcation diagram and potential energy plot of periodic
%   solutions of the von~K\'arm\'an plate model under in-plane forces
%   with $k_{2}=2$ and $k_{3}=0.5$ (even solution with $m=6$, $n=2$ in
%   \S\ref{sec:galerkin}).  The transverse load, $P_{y}$, is held
%   constant at 7. (b) Numerical solution near bifurcation point.
%   Deformation in the range $(-0.17, 0.2)$, (c) Solution at coalescent
%   limit point.  Deformation in the range $(-0.42,0)$.}
% \label{fig:bif4}
% \end{figure}

% \begin{figure}
% \includegraphics[width=\textwidth]{figure9}
% \caption{Bifurcation diagram and potential energy plot of periodic
%   solutions of the von~K\'arm\'an plate model under in-plane forces
%   with $k_{2}=2$ and $k_{3}=0.5$ (even solution with $m=6$, $n=3$ in
%   \S\ref{sec:galerkin}).  The transverse load, $P_{y}$, is held
%   constant at 7. (b) Numerical solution near bifurcation point.
%   Deformation in the range $\pm0.2$, (c) Solution at coalescent
%   limit point.  Deformation in the range $(-0.42,0)$.}
% \label{fig:bif5}
% \end{figure}

\subsubsection{Transverse stability of the striped solutions}
\label{sec:trans}
It is evident that the same striped solution exists at the limit point
of several solutions bifurcating from the flat state (e.g.\
Figures~\ref{fig:num7}(e) and \ref{fig:bif3}(c) where three distinct
two-dimensional bifurcating patterns share the same striped pattern or
Figure~\ref{fig:bif4}(c) for two others for a different value of
$P_{y}$).  The form of these sets of one-dimensional solutions are
dictated solely by the value of the load parameter $P_{y}$ only
because the end shortening in the $x$ direction, $\Delta_{x}$, is zero
and so there is no contribution to the potential energy from $P_{x}$.
On the other hand the form of the solutions emerging from the trivial
state is dependent on both $P_{x}$ and $P_{y}$.  So, the question that
inevitably arises is: how can the same striped solution be linked to
multiple two-dimensionally periodic forms?  The answer is guided both
by the approximate analysis carried out in \S\ref{sec:stab} and the
numerical simulations.

Theoretical analysis reveals that the striped solution
remains dynamically stable as long as
\begin{equation}
  P_{x}<\frac{2N_{2}^{2}}{N_{1}^{2}}.
\end{equation}
If $P_{x}$ is larger than this limit then $\eta_{1}$ (\ref{eq:eta1})
is real and the solution will grow exponentially in time.  The striped
solution under consideration closely resembles mode~1 in
Figure~\ref{fig:modes} and it is determined that the critical value of
$P_{x}\approx19.7$.  This compares reasonably well with the
numerically computed limit values of $N_{1}$ and $N_{2}$ for the full
equations which lead to the prediction $P_{x}\approx22.7$.

\begin{figure}
  \centering
  \includegraphics[width=\textwidth]{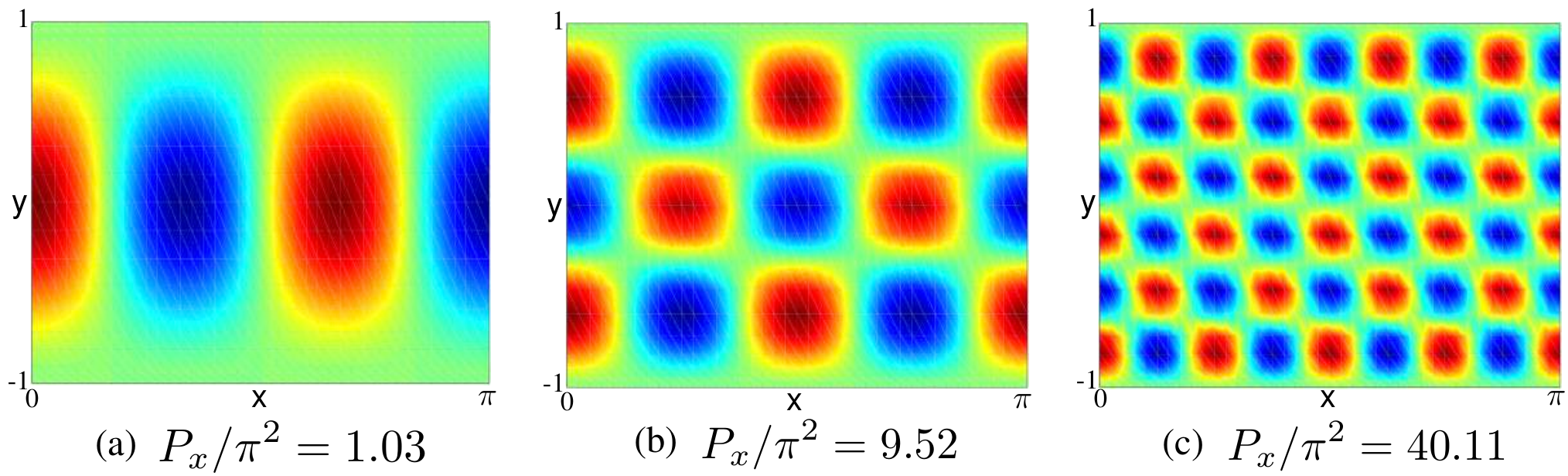}
  \caption{Eigenmodes of the striped solution depicted in
    Figure~\ref{fig:num7}(e) for various critical values of $P_{x}$
    (with $P_{y}=2$) for the cases (a)~$P_{x}=10.17$,
    (b)~$P_{x}=93.96$ and (c)~$P_{x}=395.9$.  Plots shown in the
    domain $x\in[0,\pi]$ and $y\in[-1,1]$.}
\label{fig:eig2}
\end{figure}

By numerically finding the striped solution for various value of
$P_{x}$, bifurcations can be detected by numerically estimating when
an eigenvalue of the approximate solution becomes zero.  The structure
of the eigensolution closely resembles the deflection pattern in the
vicinity of critical buckling and a bifurcation is detected at around
$P_{x}=10.17$.  The unstable eigensolution at this load is plotted in
Figure~\ref{fig:eig2}(a) and it can be seen that although the actual
deflected shape is one dimensional, the eigensolution reveals the
direction in which any disturbance would grow.  Thus we conclude that
in the vicinity of this load, the solution would naturally evolve
towards a pattern indicated by that unstable solution.  Further
transverse instabilities can be detected at various values of $P_{x}$
(while holding $P_{y}$ constant) and some are shown in
Figure~\ref{fig:eig2}~(b) and (c).  These unstable eigensolutions
resemble the buckle patterns in Figures~\ref{fig:bif3}.  The
numerically determined values of $P_{x}$ correlate very well with the
values at the limit points of the bifurcation curves.  Thus we
conclude that the same striped pattern can indeed be linked to various
two-dimensional patterns in a systematic way.

% \begin{figure}
%   \centering
%   \includegraphics{unst-eigen-A}
%   \caption{Eigensolution of the striped deflection pattern depicted in
%     Figure~\ref{fig:num7} with $P_{x}=29.21$.}
%   \label{fig:eig1}
% \end{figure}

We have found that for each given $P_{y}$, there is an associated
striped solution.  However, we can subsequently induce buckle patterns
of different types by then loading the plate in the orthogonal
direction, $P_{x}$.  Theoretically, this could be used as a way of
producing desired deflection pattern in a repeatable manner.  A
consequence of the phenomenon of multiple patterns being able to
develop from a particular 1D striped mode could be used to manufacture
a layer of material with a desired pattern in a two-step process.
Firstly, the specimen could be loaded in one in-plane direction to
give a striped buckle shape and then judicious loading in the
orthogonal direction would then give rise to the required 2D pattern.
In particular, carefully varying the secondary load would, in
principle, be a possible method of manufacturing materials with
various tuneable patterns and hence physical properties \cite{Yang15}.

\subsection{Localized behaviour}
As we indicated in the previous section, periodicity is the
predominant form of behaviour to be expected in most plated structures
since the finite domain precludes genuinely localized solutions which
are only evident over an infinite domain.  However, if the domain is
significantly larger than the natural wavelength of the buckle pattern
then a (periodic) solution can arise which comes to resemble its truly
localized counterpart in the sense that over the whole extent of the
plate, only a small fraction undergoes noticeable deflection.
Localization of patterns can be favourable under certain conditions,
particularly where a softening nonlinearity makes such a pattern
energetically more favourable to jump to than its periodic counterpart
\cite{HBT} even though both are kinematically admissible.  Even in the
case that such localized solutions are unstable, they can form
critical (mountain pass) modes \cite{HoLoPe} to other forms. Thus a
localized solution may develop in a structure of finite extent if at
one end the solution and all its derivatives approach zero.  Such is
the case for the buckling of a thin cylindrical shell \cite{HuLoCh}.
In other applications, such as contemporary material science, periodic
buckling has been the primary focus \cite{ZhSoYa}.  However, if the
extent of the domain is much larger than the wavelength of buckle
patterns then isolated (localized) patterns can coexist with periodic
solutions.

\begin{figure}
  \centering
  \includegraphics[width=\textwidth]{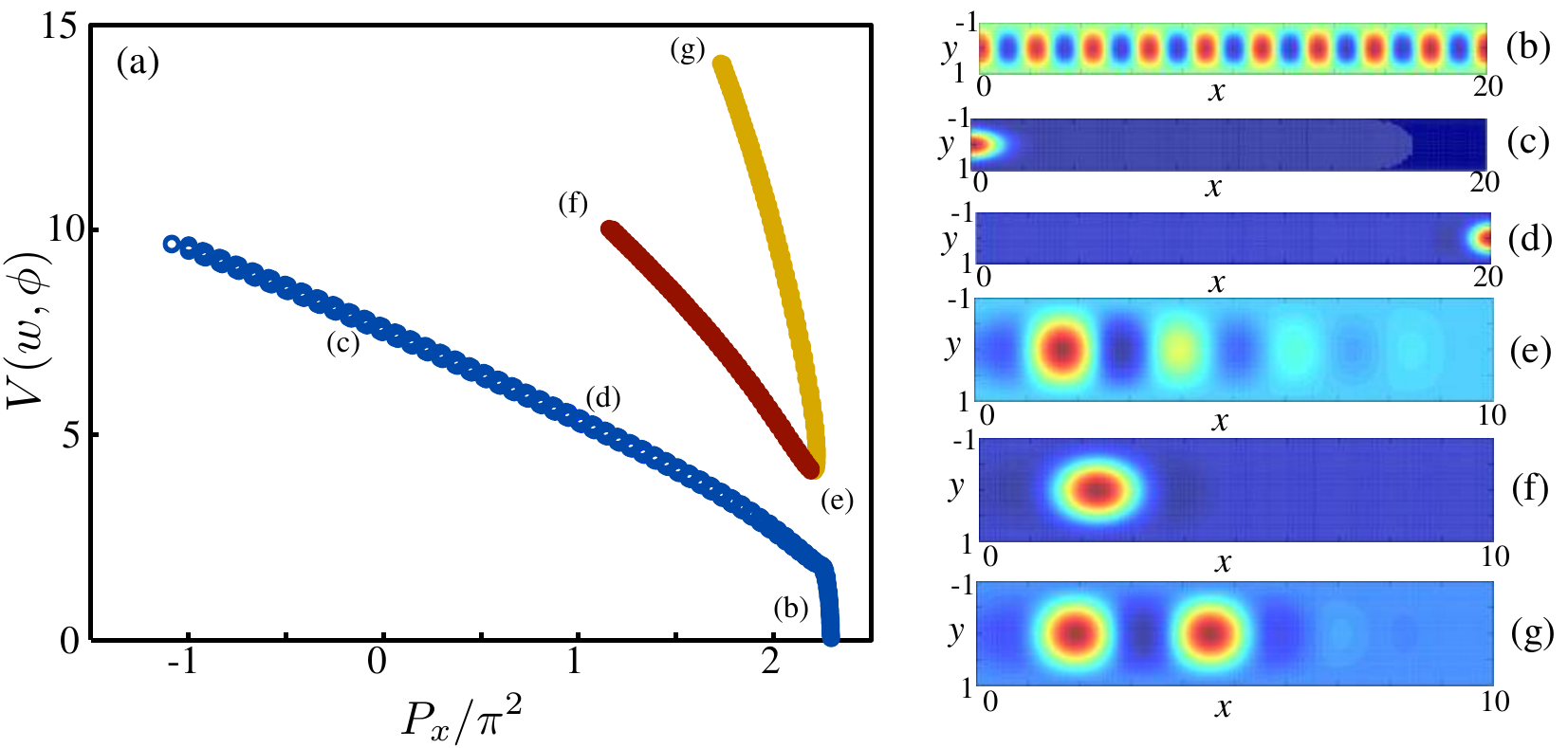}
  \caption{Localized solutions of the plate.  Two families are shown
    in the bifurcation diagram (a): a primary solution calculated over
    $x\in[0,20]$ (blue) bifurcating from the trivial state with
    deflections shown in (b)~near the bifurcation point itself
    ($w\in[-7\times10^{-3},7.1\times10^{3}]$); (c)~and (d)~further
    down the two branches (seen as a single line as the overlap)
    ($w\in[0,0.62$); and a secondary solution calculated over
    $x\in[0,10]$ with deflections shown at the limit point
    (e)~($w\in[-0.11,0.24]$) at the end of each of the branches
    (f)~and (g)~(shown as red (dark) and yellow (light) respectively)
    and $w\in[-0.03,0.53]$ and $w\in[-0.12,0.42]$, respectively.}
  \label{fig:loc1}
\end{figure}

To demonstrate the relationship between periodic and localized forms,
Figure~\ref{fig:loc1} shows the bifurcation structure of two plates of
non-dimensional lengths 10 and 20.  These plots can be interpreted as
a localized patterns with symmetry about the postion $x=0$.  As can be
seen, a pair of solutions bifurcate from the trivial (zero-energy)
state, the other is a subsidiary solution forming a multi-humped
pattern \cite{WaBa}.  We infer that as length increases the limiting
behaviour will be localized states that bifurcate from the unbuckled
state and co-exist with their periodic counterparts just as in the
one-dimensional strut model \cite{HBT}.  For a finite-length plate,
the way in which a localized pattern may be realized is by an initial
periodic pattern which is then disturbed, perhaps by a secondary
bifurcation, into a localized state.  Such multi-staged buckling
phenomena are common in plates \cite{HdS,HuAhm}.

\subsection{Aspect ratio}
\label{sec:aspect}
\begin{figure}
  \centering
  \includegraphics[width=\textwidth]{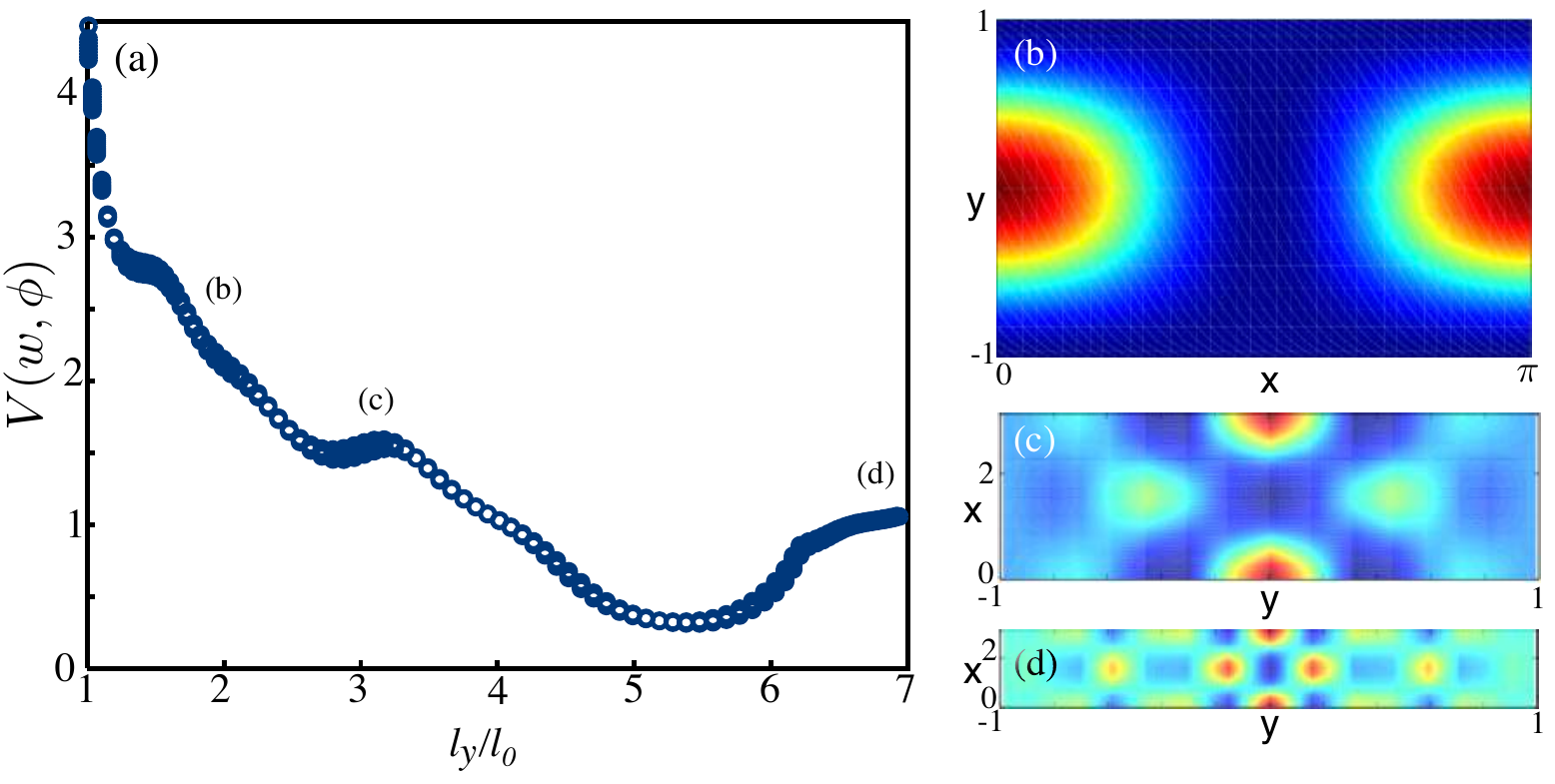}
  \caption{Aspect ratio variation for $P_{x}=19.74$, $P_{y}=2$,
    $k_{2}=2$ and $k_{3}=0.5$.  (a)~Variation of total potential
    energy, $V$ against aspect ratio. (b)~Solution with original
    aspect ratio $l_{0}=2/\pi$ ($w\in[-0.10,0.32]$).  (c)~Continued
    solution with $l_y=3.2468l_{0}$ ($w\in[-0.05,0.14]$) and
    (d)~solution with $l_y=6.9393l_{0}$ ($w\in[-0.04,0.06]$).  Note
    that plots (c) and (d) have been rotated through $90^\circ$.}
  \label{fig:aspect}
\end{figure}

The problem as formulated in this study consists of an investigation
into the theoretical buckling aspects of a structure that extends into
two dimensions.  The equivalent one-dimensional problem (the strut
model) has been extensively studied but the plate on a foundation has
attracted far less interest.  There are many potential avenues to
explore for this model and we have hinted at some possible ideas here.
In many ways, it is perhaps the role played by the aspect ratio that
is perhaps most fundamental so we introduce the aspect ratio,
$l_{y}\equiv\gamma/L$, as a continuous parameter.  In
Figure~\ref{fig:aspect} we illustrate the evolution of a basic
periodic solution as $l_y$ is varied from its original value
$l_{0}=2/\pi$. As the ratio is increased so the pattern spreads to
fill the extra width and yet the buckled cells approximately maintain
their own aspect ratio.  As the width gets larger, there are a few
transitions when the number of cells increases by one each time; this
is clearly a form of mode jumping \cite{EvHu99}.  Further evidence
that lends credence to this conclusion is provided in the plot of
potential energy versus $l_{y}$ as the curve exhibits undulations at
the point where the number of buckle cells changes.

\section{Conclusions and further study}
\label{sec:conc}
The current work has examined the fundamental behaviour of a
structural system in order to investigate the effect of adding an
extra spatial dimension to a well-studied one-dimensional problem.  As
expected, the introduction of a finite width opens up a new richness
of possibilities of solutions available after buckling ensues.  The
loading regime is seen to be enhanced with the ability to apply
different forces in different directions and indeed to allow the
loading to vary spatially, although that has been left for future
work.

We have seen that some conventional techniques, such as the Galerkin
method, are able approximately to capture buckling behaviour emergent
from the flat state but fully numerical methods are needed to follow
what happens to these solutions further into post-buckling.  We found
that there are special one-dimensional (striped) patterns which
eventually connect both the branches which emerge from the critical
state and furthermore that the same striped pattern (governed by the
transverse load, $P_{y}$) can lie at the end of multiple branches.
The unstable eigensolutions of the striped solution reveal which
two-dimensional (spot) solution it links to.  An interesting
possibility arises from our findings in that we have outlined a
plausible procedure by which material could be manufactured with
pre-determined characteristics by careful application of load in
one in-plane direction followed by loading at right angles to it.
This technique could be a simpler alternative to current methods used
to manufacture highly-tuned nanomaterials.

A straightforward analysis of the striped solution using an adaptation
of the strut model \cite{WHW} reveals information about the stability
of solutions and this, coupled to numerics, has enabled us to locate
the boundary between stable and unstable behaviour.  The boundary
gives a reasonable approximation for the longitudinal load at which the
striped solution gives way to spots or cellular buckle patterns.

Although we have considered the loading as two independent parameters
this study, it may be instructive in future to allow the load to vary
along either or both edges in order to simulate various conditions
that may arise in reality.  An example of such varying stress may
arise through thermal expansion or contraction of a plate embedded in
an elastic medium.  Some consideration has been given to localized
behaviour and the effects of aspect ratio.  These features would
increase in importance were we to look at the case where the buckling
cells are small compared with the length and width of the plate.

The plate--foundation model has been used to mimic many physical
situations ranging from material wrinkling in industrial manufacturing
\cite{DoBuHu14}, theoretical neo-Hookian materials \cite{CaoHu11} and
fold hierarchies in thin films \cite{Reis11} through to the
morphological development of tissue and organs such as the brain
\cite{Bud15}.  However, most modelling usually starts with some
simplifying assumptions including the reduction of the problem into
one-dimensional phenomena.  We offer here a way to begin
systematically to study two-dimensional behaviour which holds the
possibility of uncovering a much richer set of competing and
interacting solutions. Although resolution is presently far off,
progress in this direction may conceivably go some way to explain some
of the complex behaviour observed in practice.

{\bf Ethics}: This work did not involve any experiments on living
  organisms.

{\bf Data Accessibility}: Calculations were performed using equations which are all
  published in this paper and numerical computations were carried out
  using published algorithms with
  \textsc{Matlab}.% There are no physical experimental
% data.

{\bf Funding}: The research was funded as part of the authors' standard
  employment contracts.

{\bf Authors' Contributions}: MKW conceived of the problem, MKW and DJBL designed the
  programme of study and MKW undertook the Galerkin calculations and
  computations based on numerical algorithms developed by DJBL.  MKW
  and APB worked on the reduced one-dimensional stability problem and
  DJBL devised a numerical verification method for these results.  All
  authors contributed to the writing of the paper and approved its
  final submission.

{\bf Competing Interests}: The authors had no competing interests in performing this
  work.

{\bf Acknowledgements}: We express our thanks to Professor Geoffrey Nash at the
  University of Exeter for helpful discussions regarding the
  manufacture and properties of modern two-dimensional materials and
  undergraduate David Frampton, also at the University of Exeter, for
  helping to develop the \textsc{Matlab} code to find the Galerkin
  approximations in \S\ref{sec:galerkin}.  Finally, we are grateful to
  the referees for their helpful comments which considerably enhanced
  the presentation of this work.

\bibliographystyle{siam}
\bibliography{references}

\end{document}